\documentclass[letterpaper, 10 pt, conference]{ieeeconf}  

\IEEEoverridecommandlockouts                        
\usepackage{amsmath}
\usepackage{amsfonts}
\usepackage{graphicx}
\usepackage{algorithm}
\usepackage{caption}
\usepackage{subcaption}
\usepackage[noend]{algpseudocode}
\usepackage{cite}

\newcommand{\tr}{\mathrm{tr}}

\newcommand{\dq}{\,\mathrm{dq}}
\newcommand{\dtprime}{\,\mathrm{dt'}}

\newcommand{\norm}[1]{\left\lVert#1\right\rVert}

\newtheorem{proposition}{Proposition}

\newtheorem{assumption}{Assumption}

\def\Vec{\mathop{vec}\nolimits}

\title{\LARGE \bf
Optimal Periodic Multi-Agent Persistent Monitoring of a Finite Set of Targets with Uncertain States
}

\author{Samuel C. Pinto$^1$, Sean B. Andersson$^{1,2}$, Julien M. Hendrickx$^3$, and Christos G. Cassandras$^{2,4}$
\\
$^1$Dept. of Mechanical Engineering, $^2$Division of Systems Engineering,\\ $^4$Dept. of Electrical and Computer Engineering \\
Boston University, Boston, MA 02215, USA \\
$^3$ICTEAM Institute, UCLouvain, Louvain-la-Neuve 1348, Belgium \\
\{samcerq,sanderss,cgc\}@bu.edu, julien.hendrickx@uclouvain.be
        \thanks{This work was supported in part by NSF under grants ECCS-1509084, ECCS-1931600, DMS-1664644, CNS-1645681, and CMMI-1562031, by ARPA-E's NEXTCAR program under grant DE-AR0000796, by AFOSR under grant FA9550-19-1-0158,  and by the MathWorks. The work of J. Hendrickx was supported in part by Communat\'{e} fran\c{c}aise de Belgique - Actions de Recherche Concert\'{e}es and a WBI World Excellence Fellowship.
}
}

\begin{document}

\maketitle
\thispagestyle{empty}
\pagestyle{empty}

\begin{abstract}
We investigate the problem of persistently monitoring a finite set of targets with internal states that evolve with linear stochastic dynamics using a finite set of mobile agents. We approach the problem from the infinite-horizon perspective, looking for periodic movement schedules for the agents. Under linear dynamics and some standard assumptions on the noise distribution, the optimal estimator is a Kalman-Bucy filter and the mean estimation error is a function of its covariance matrix, which evolves as a differential Riccati equation. It is shown that when the agents are constrained to move only over a line and they can see at most one target at a time, the movement policy that minimizes the mean estimation error over time is such that the agent is always either moving with maximum speed or dwelling at a fixed position. This type of trajectory can be fully defined by a finite set of parameters. For periodic trajectories, under some observability conditions, the estimation error converges to a steady state condition and the stochastic gradient estimate of the cost with respect to the trajectory parameters of each agent and the global period can be explicitly computed using Infinitesimal Perturbation Analysis. A gradient-descent approach is used to compute locally optimal parameters. This approach allows us to deal with a very long persistent monitoring horizon using a small number of parameters.
\end{abstract}

\section{INTRODUCTION}
\label{sec:intro}

As autonomous cyber-physical systems are continuously increasing their importance in our society, the topic of long term autonomy is gaining more interest. In the context of long term autonomy one is not looking only to accomplish short term goals, but also to plan behaviors that will be efficient in the long term. One class of problems of interest in the context of long term autonomy is where one has a collection of points of interest (denoted as "targets") and a set of moving agents that can visit these targets and perform some form of estimation or control to their internal state. This paradigm finds applications in very diverse contexts, such as traffic surveillance in critical points of a city, sea temperature estimation and tracking of macro particles in optical microscopy. While for static systems the estimation or control error does not grow over time, in general dynamic and stochastic systems this error may grow very fast as time increases. Therefore, if there are not enough agents to continuously estimate or control these targets, then the mobile agents must travel over the environment with trajectories which can persistently visit the targets in order to avoid unbounded errors as time goes to infinity. Persistent monitoring is the term used to refer to this class of problems.

While the persistent monitoring problem has already been studied in the literature \cite{Stump:2011gv,Yu:2016fn,Yu:2017iw,Yu:2018cf,cassandras2013optimal,zhou2018optimal,lan2013planning,lan2014variational}, these works focused on analyzing the transient behavior of the system. Motivated by the prospects of long term autonomy, we tackle the problem from the infinite horizon point of view, where continuous estimation of internal states of the targets is performed. For a periodic solution, the mean estimation error of these internal states, will, as time goes to infinity, approach the estimation error of a steady state periodic solution independent of the initial conditions. While the idea of periodicity of the solution of the persistent monitoring problem has already been explored in \cite{lan2013planning,lan2014variational}, these works did not provide tools for analyzing the behavior of the solution in steady state. Therefore, in order to apply these techniques for in long term one would either need to optimize over a very long period or always recompute the solution for the next cycle and both approaches have an expressive computational overhead. As a way to overcome this issue, instead of minimizing the transient estimation error, we can neglect the transient effects and plan trajectories that minimize the steady state estimation error of a periodic trajectory. In this paradigm, it is only necessary to optimize the parameters that describe one period of the trajectory, which is usually a very small number of parameters. Moreover, as time goes to infinity, the mean estimation error will be arbitrarily close to the state state error that has been planned for.

In this work, we provide tools for analyzing and optimizing a periodic trajectory in order to minimize the steady state estimation error. We assume that agents can observe the targets' internal states with a linear observation model with Gaussian additive noise, and hence, the optimal estimator for this model is a Kalman-Bucy filter and the differential Riccati equation expresses the dynamics of the covariance matrix and, naturally, the mean quadratic estimation error. We extend the work \cite{pinto2019monitoring}, in which we considered targets distributed in a 1-D environment and where the agent can see at most one target at a time, for in this scenario we are able to show that there is a parameterization of the optimal solution of the finite-time version of the problem considered here. In this paper, however, we consider the infinite horizon version and restrict ourselves to periodic trajectories for which we show that, under some assumptions, the covariance matrix converges to a limit cycle. We then use Infinitesimal Perturbation Analysis (IPA) in a centralized gradient descent scheme to obtain locally optimal trajectories. This approach not only allows the shape of the trajectory to be optimized, but also its period. 
It is worth noticing that in many interesting applications that can be modeled as a persistent monitoring problem, agents are constrained to (possibly multiple) uni-dimensional mobility, such as powerline inspection agents, cars on streets, and autonomous vehicles in rivers.

The remainder of this paper is organized as follows. Sec. \ref{sec:formulation} presents the problem formulation, including target and agent dynamics and the Kalman-Bucy filter for estimating target states from the agent measurements. Sec. \ref{sec:opt_control} discusses properties of the optimal control, leading to a parameterized representation of an optimal trajectory. In Sec. \ref{sec:steady_state}, properties and conditions for convergence of the Riccati equation to a limit cycle solution are discussed.  The IPA-driven gradient descent is considered in Sec. \ref{sec:hybrid_sys} and the entire scheme is demonstrated through simulations in Sec. \ref{sec:results}. Finally, Sec. \ref{sec:conclusion} gives a conclusion and shares ideas for future works.

\section{PROBLEM FORMULATION}
\label{sec:formulation}
We consider an environment with $M$ fixed targets located at positions $x_1,...,x_M\in \mathbb{R}$. Each target has an internal state ${\phi_i} \in \mathbb{R}^{L_i}$ with dynamics
\begin{equation}
    \label{eq:dynamics_phi}
    {\dot{\phi}}_i(t) = A_i{\phi}_i(t) + {w}_i(t),
\end{equation}
where $w_i,$ $i=1,\dots, M,$ are mutually independent, zero mean, white, Gaussian distributed processes with $E[{w}_i(t){w}_i(t)^T]=Q_i$ with $Q_i$ a positive definite matrix for every $i$.

We have $N$ mobile agents, whose positions at time $t$ are denoted by $s_1(t),...,s_N(t)\in\mathbb{R}$, equipped with sensing capabilities. These agents can move with the following the kinematic model
\begin{equation}
    \dot{s}_j(t)=u_j(t),\ j=1,...,N, \label{eq:dynamics_agents}
\end{equation}
where their speed is constrained by $|u_j(t)| \leq 1$, after proper scaling. Note that, even though we only consider first order dynamics in this paper, extensions to second order dynamics would likely follow similar results, as discussed in \cite{wang2019}. The internal state of target $i$ can be observed by agent $j$ according to the following linear model.
\begin{equation}
    \label{eq:observation_model_ij}
    {z}_{i,j}(t)=\gamma_j\left(s_j(t)-x_i\right)H_i{\phi}_i(t)+{v}_{i,j}(t),
\end{equation}
where ${v}_{i,j}$, $i=1,\dots,M$, $j=1,\dots, N$ are mutually independent zero mean, white, Gaussian distributed noise processes, independent of the ${w}_i$, with $E[{v}_{i,j}(t) {v}^T_{i,j}(t)]=R_i$, $R_i$ positive definite, and $\gamma_j(\cdot)$ is a scalar function. In this model, the noise power is constant but sensed the signal level varies as a function of the distance to the target. Even though the analysis conducted in this paper is valid for any unimodal $\gamma_j(\cdot)$ that has finite support, we use the following definition for concreteness:
\begin{equation}
    \label{eq:model_gamma}
    \gamma_j(\alpha) =    
    \begin{cases}
        0, & |\alpha|>r_j,\\
        \sqrt{1-\frac{|\alpha|}{r_j}}, & |\alpha|\leq r_j.
    \end{cases}
\end{equation}

\begin{figure}[htp!]
    \centering
    \includegraphics[width=0.4\textwidth]{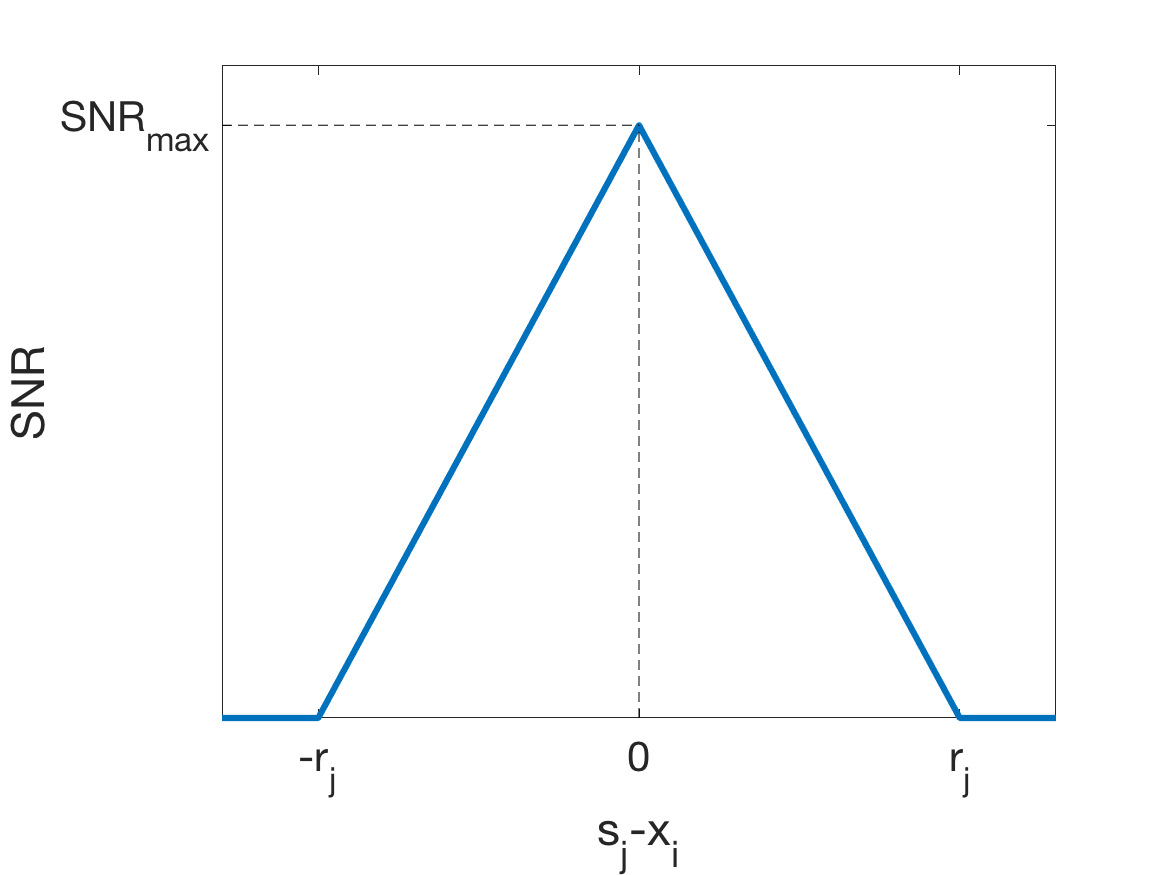}    
    \caption{Illustration of the dependence of the SNR on the distance between agent $j$ and target $i$.}
    \label{fig:snr}
\end{figure}

Under this model, the instantaneous signal to noise ratio (SNR) of a single measurement made by agent $j$ is given by
\begin{multline}
    \label{eq:snr}
    \frac{E\left[({z}_{i,j}(t)-{v}_{i,j}(t))^T({z}_{i,j}(t)-{v}_{i,j}(t))\right]}{E[{v}_{i,j}^T(t){v}_{i,j}(t)]} \\ =\max \left(0,1-\frac{|s_j-x_i|}{r_j}\right)\frac{{\phi}_i^T(t)H_i^TH_i{\phi}_i(t)}{\text{tr}(R_i)},
\end{multline}
where $\tr(\cdot)$ is the trace of a matrix. 
The term ${\phi}_i^T(t)H_i^TH_i{\phi}_i(t)(\text{tr}(R_i))^{-1}$ is deterministic and scalar and can not be influenced by the relative position between the agent and the target. On the other hand, the max function (along with the SNR) is maximum when the agent's position coincides with that of the target, linearly decreases as it moves farther, and is zero if the distance is greater than $r_j$. Therefore, useful information can only be acquired within the sensing range of the agent and within this range the measurement quality is higher the closer the agent is to the target. Figure \ref{fig:snr} illustrates this dependence.

The instantaneous joint observations performed by all the agents of the same target can be written as a vector of observations,
\begin{equation}
    \label{eq:vector_observation_model}
    {z}_i(t) = [{z}_{i,1}^T,...,{z}_{i,N}^T]^T = \tilde{H}_i(s_1,...,s_n){\phi}_i(t)+\tilde{{v}}_i(t)
\end{equation}
where
\begin{align}
    \label{eq:def_h_tilde}
    \tilde{H}_i&=[\gamma_1(s_1-x_i)H_i^T,\cdots,\gamma_N(s_N-x_i)H_i^T]^T, \\
    \tilde{{v}}_i(t)&=[{v}_{i,1}^T(t),...,{v}_{i,N}^T(t)]^T,
    \label{eq:def_r_tilde}
 \end{align}
 and
 \begin{align}
    E[\tilde{ v}_i^T(t)\tilde{{ v}}_i(t)]&=\tilde{R}_i=
    \begin{bmatrix} 
    {R}_{i} & {0} & \dots & {0}\\
    {0} & R_i & \dots & {0}\\
    \vdots & \vdots  &\ddots & \vdots \\
    {0}& {0} & \dots   &     R_i 
    \end{bmatrix}.
\end{align}

Note that \eqref{eq:dynamics_phi} and \eqref{eq:vector_observation_model} define a linear, time-varying, stochastic system, if the trajectories are already pre-defined. The optimal estimator for the states ${\phi}_i(t)$ is then a Kalman-Bucy Filter \cite{bucy1968filtering}. A proof that this is indeed the optimal estimator is not ommited here for space reasons, but the derivation is analogous to the similar result in \cite{lan2014variational}, where it is shown that the Kalman-Bucy filter is an optimal estimator, considering targets with internal states with the same dynamics as in \eqref{eq:dynamics_phi} and a general agent dependent time-varying observation model, similar to \eqref{eq:observation_model_ij}.

We denote $\hat{{\phi}}_i(t)$ the estimate of the current state of ${\phi}_i(t)$, and ${e_i}(t)=\hat{{\phi}}_i(t)-E[\hat{{\phi}}_i(t)]$ the estimation error and $\Omega_i=E[{e}_i(t){e}_i^T(t)]$  the error covariance matrix. Then, the Kalman-Bucy filter equations are 
\begin{subequations}
	\begin{align}
    \dot{\hat{{\phi}}}_i(t)&=A_i\hat{{\phi}}_i(t)+\Omega(t)_i\tilde{H}_i^T(t)\tilde{R}_i^{-1}\left(\tilde{{z}}_i(t)-\tilde{H}_i(t)\hat{{\phi}}_i(t)\right), \\
    \dot{\Omega}_i(t) &= A_i\Omega_i(t)+\Omega_i(t)A_i^T+Q_i-\Omega_i(t)\tilde{H}_i^T\tilde{R}^{-1}_i\tilde{H}_i\Omega_i(t). \label{eq:dynamics_omega_matricial}   
	\end{align}
\end{subequations}

Substituting \eqref{eq:model_gamma}, \eqref{eq:def_h_tilde}, and \eqref{eq:def_r_tilde} into \eqref{eq:dynamics_omega_matricial} yields
\begin{multline}
    \label{eq:dynamics_omega}
    \dot{\Omega}_i(t) = A_i\Omega_i(t)+\Omega_i(t)A_i^T+Q_i\\
    -\Omega_i(t)G_i\Omega_i(t)\eta_i(t),
\end{multline}
where $G_i=H_i^TR_i^{-1}H_i$ and
\begin{equation}
    \label{eq:def_eta}
    \eta_i(t)=\sum_{j\in C_i(t)}\left(1-\frac{|s_j(t)-x_i|}{r_j}\right).    
\end{equation}
$C_i(t)$ is the \textit{agent neighborhood} of target $i$, i.e., the indices of all agents with target $i$ within their respective sensing range at time $t$.

The overall goal is to minimize the mean estimation error over an infinite time horizon. Formally, for the set of inputs $u$ where the following limit exists, the objective is to find the optimal cost $J^\star$ (where the input dependence on the time is ommited for the sake of notation conciseness):
\begin{equation}
    \label{eq:expectation_opt_objective}
    J^\star=\min\limits_{u_1,...,u_N}\lim_{t\rightarrow \infty}\frac{1}{{t}}\int_0^{t}\left(\sum_{i=1}^M E\left[{e}_i^T(t'){e}_i(t')\right]\right) \dtprime.
\end{equation}
Using the fact that 
\begin{align*}
	E\left[{e}_i^T(t){e}_i(t)\right]=\text{tr}(E\left[{e}_i(t){e}^T_i(t)\right])=\tr(\Omega_i)	
\end{align*}
the optimization in \eqref{eq:expectation_opt_objective} can be rewritten as
\begin{equation}
    \label{eq:opt_objective}
    \min\limits_{u_1,...,u_N}J=\lim_{t\rightarrow \infty}\frac{1}{t}\int_0^{t}\left(\sum_{i=1}^M\text{tr}\left(\Omega_i(t')\right)\right) \dtprime,
\end{equation}
subject to the dynamics in \eqref{eq:dynamics_agents} and \eqref{eq:dynamics_omega}.

\section{OPTIMAL CONTROL PROPERTIES}
\label{sec:opt_control}
The purpose of this section is to establish properties of the optimal control solution in order to be able to describe the optimal trajectory by a finite set of parameters. In previous work \cite{pinto2019monitoring}, we used Hamiltonian Analysis to derive these properties for non-periodic trajectories. However, the same argument cannot be used for periodic trajectory since the latter derivation was deeply dependent on the fact that the terminal position of the agent was not specified, which gave a terminal condition for the optimal costate matrix associated to the covariance. Therefore, we establish a similar results that also contemplates periodic trajectories, where we constraint ourselves to trajectories where the terminal and the initial point coincide over the course of one period. Initially, we introduce the following proposition which is going to be essential in the proof of the properties of an optimal control solution.

\begin{proposition}
\label{prop:monotonicity_ricatti_eq}
 Given $\Omega_1(t)$ and $\Omega_2(t)$, two bounded covariance matrices under the dynamics in \eqref{eq:dynamics_omega} with $A=A_1=A_2$, $G=G_1=G_2$, $Q=Q_1=Q_2$, then if $\Omega_1(0)-\Omega_2(0)$ is negative semi-definite and $\eta_1(t) \geq \eta_2(t)$, then $\Omega_1(t)-\Omega_2(t)$ is a negative semi definite matrix for $t\geq 0$.
\end{proposition}
\begin{proof}
    Defining $\Xi=\Omega_1(t)-\Omega_2(t)$. The dynamics of $\Xi$ is described by the following equation.
    \begin{multline}
        \label{eq:dynamics_xi_initial}
        \dot{\Xi}(t) = A\Xi(t)+\Xi A^T-\eta_1(t)\Omega_1(t)G\Omega_1(t)\\+\eta_2(t)\Omega_2(t)G\Omega_2(t).
    \end{multline}
    Adding and subtracting the terms $\eta_1(t)\Omega_2(t)G\Omega_2(t)$ and $\eta_1(t)\Omega_1(t)G\Omega_2(t)$ to the equation, we can rewrite the relation in \eqref{eq:dynamics_xi_initial} as:
    \begin{multline}
        \label{eq:dynamics_xi}
        \dot{\Xi}(t) = A\Xi(t)+\Xi A^T-\eta_1(t)\left[\Omega_1(t)G\Xi(t)+\Xi(t)G\Omega_2(t)\right]\\+\left[\eta_2(t)-\eta_1(t)\right]\Omega_2(t)G\Omega_2(t).
    \end{multline}
    From Thm.~1.e in \cite{kriegl2011denjoy}, since $\Xi(t)$ is a $C^1$ matrix, its eigenvalues 
can be $C^1$ time parameterized. Let $\mu_n$ denote the $n^{th}$ eigenvalue of $\Xi(t)$ and $x_n(t)$ the corresponding unit norm eigenvector. 
Then, from Thm.~5 in \cite{lancaster1964eigenvalues} we have that
    \begin{equation*}
        \dot{\mu}_n=  x_n^T\dot{\Xi} x_n.
    \end{equation*}
Also, notice that by using \eqref{eq:dynamics_xi} and the fact that $\lambda_{\min}\left(\frac{D+D^T}{2}\right)\leq x^TDx\leq \lambda_{\max}\left(\frac{D+D^T}{2}\right)\norm{x}=\norm{D}\norm{x}$, for any square matrix D,
\begin{align*}       
    \dot{\mu}_n &\leq \norm{A}\mu_n-\eta_1\xi\mu_n+\left[\eta_2-\eta_1\right]x_n^T\Omega_2G\Omega_2x_n \\
    &\leq \norm{A}\mu_n-\eta_1\xi\mu_n
\end{align*}
where $\xi=\lambda_{min}\left((\Omega_1+\Omega_2)G+G(\Omega_1+\Omega_2)\right)$.
Using Gronwall's inequality and the fact that a first order linear ODE does not change sign, we conclude that $\mu_n(t)\leq0, \forall\ t\in [0,T]$ and, therefore, $\Xi(t)$ is negative semidefinite.
\end{proof}
In Prop. \ref{prop:monotonicity_ricatti_eq}, even though it does not fully match the notation established in Sec. \ref{sec:formulation}, $\Omega_1$ and $\Omega_2$ can also be understood as covariance matrices for the same target but under different agent trajectories and this is the way that they are going to be interpreted in Prop. \ref{prop:optimal_control}. 

Before proceeding to the proposition about an optimal control structure, a few definitions are necessary. An isolated target $i$ is a target such that 
\begin{align*}
	\min\limits_{k\neq i}|x_i-x_k|>2r_{\max}, \quad r_{\max}=\max\{r_1,...,r_N\}.
\end{align*} and the minimum distance between visible areas ($d_{\min}$) is defined as:
\begin{align*}
	d_{\min} = \min_{i,k}|x_i-x_k|-2r_{\max}>0.
\end{align*}
and the finite time cost is defined as
\begin{equation}
    \label{eq:finite_time_opt_objective}
    J(u_1,...,u_N)=\frac{1}{t}\int_0^{t}\left(\sum_{i=1}^M\text{tr}\left(\Omega_i(t')\right)\right) \dtprime.
\end{equation}

We can, then, claim a similar result to Prop. 1 in \cite{pinto2019monitoring}.

\begin{proposition}
    \label{prop:optimal_control}
    In an environment where all the targets are isolated, given any policy $u_j(t')$, $j=1,...,N$, then there is a policy $\tilde{u}_j(t')$ where $\tilde{u}_j(t')\in\{-1,0,1\}$ where 
    $J(u_1,...,u_N)\geq J(\tilde{u}_1,...,\tilde{u}_N)$ and the number of control switches is upper bounded by $2\frac{t}{d_{\min}}+4$.
\end{proposition}
\begin{proof}
    This proposition is going to be proved by construction: given a policy $u_j(t')$ with $\eta_i(t')$ associated to it (as defined by \eqref{eq:def_eta}), we will construct an alternative policy $\tilde{u}_j(t')$ associated with $\tilde{\eta}_i(t')$ such that $\tilde{\eta}_i(t') \geq \eta_i(t')\ \forall t'\in[0,t]$ and $i=1,...,M$, and then use Prop. \ref{prop:monotonicity_ricatti_eq}, along with the definition of the cost \eqref{eq:finite_time_opt_objective}, to show that the alternative policy has lower or equal cost than the original one.
    
    Initially, we focus on the policy $u_j(t')$.  Note that an agent $j$ is said to visit a target $i$ if at some time $t'$, $|s_j(t')-x_i(t')|<r_j$. For every agent in the policy $u_j(t')$, there is an ordered collection of targets it visits in $[0,t]$. Therefore, there must exist a set of indices of all the targets visited by agent $j$, $y^j_0,...,y^j_{K_j}\in \{1,...,M\}$, such that $y_p^j\neq y_{p-1}^j$ and agent $j$ visited no other target in the time between visiting targets $y_p^j$ and $y_{p-1}^j$. This set is an ordered set of all the targets that agent $j$ visited over $[0,t]$, not counting consecutive visits to the same target. Notice that the same target can be present more than once in the vector $[y^j_0,...,y^j_{K_j}]$ but if that is the case, it will not be in consecutive positions.
    
    For each of these visits, we can define the initial visitation time $t_p^j$ for $p=1,...,K_j$ as
    \begin{align*}
        t_p^j = \inf \{t' |t'>t_{p-1}^j \text{ and agent $j$ visits target $y_p^j$ at time $t'$} \}, 
    \end{align*}
    and $t_0^j=0$ and $t_{K_j+1}^j=t$
    Also, we define the initial visitation positions $a_p^j$, i.e., the position of the agents when it starts visiting a target, $a_p^j=s_j(t_p^j)$, $p=0,...,K_j+1$. $\chi_p^j$ is the position of the target $y_p^j$ when a visit starts, i.e. $\chi_p^j=x_{y_p^j}$. Also note that while $t' \in [t_{p-1}^j,t_p^j)$, agent $j$ only influences the value of $\eta_i(j)$ of the target it is currently visiting.
    
    We propose the following alternative policy, where $\tilde{u}_j(t')$ for $t'\in[t_{p-1}^j,t_{p}^j)$ is such that:
    \begin{equation}
        \label{eq:alternative_u}
        \tilde{u}_j(t')=\begin{cases}
            \frac{a_p^j-s_j(t')}{|a_p^j-s_j(t')|},\text{ if }t_p^j-t' \leq |a_p^j-s_j(t)|,\\
            0, \text{ if }t_p^j-t' > |a_p^j-s_j(t)| \text{ and }s_j(t') = \chi_p^j, \\ 
            \frac{\chi_p^j-s_j(t')}{|a_p^j-s_j(t')|},\text{ otherwise.}
        \end{cases}
    \end{equation}
    
    The intuition behind the proposed alternative policy is that at the beginning of each visit, the agent moves with maximum speed towards the target $y_p^j$ and if it reaches the target, it dwells on top of it. However, it must move in a way such that it begins the next visit at the same time that the original policy, i.e., the positions of agent $j$ associated to the alternative policy $\tilde{s}_j(t')$ is such that $\tilde{s}_j(t_p^j)={s}_j(t_p^j)=a_p^j$.
    
    Notice that the provided construction provides a feasible trajectory, since the original trajectory is assumed feasible. Also, for time $t'\in [t_{p}^j,t_{p+1}^j]$ both the original and the alternative policies only influence value of $\eta_i$ for $i=t_p^j$ and, since in the alternative policy all the agents are closer or at least as close to the currently visited target, by looking at \eqref{eq:def_eta} we can claim that
    \begin{equation*}
        \tilde{\eta}_i(t')\geq \eta_i(t'),\ \forall t'\in[0,t],\ i\in\{1,...,M\}.
    \end{equation*}
    Therefore, using Prop. \ref{prop:monotonicity_ricatti_eq} and the cost definition \eqref{eq:finite_time_opt_objective}, we get that
     \begin{multline*}
    J(\tilde{u}_1,...,\tilde{u}_N)-J({u}_1,...,{u}_N)=\\
    \frac{1}{t}\int_a^b\sum_{i=1}^M\tr{\left(\tilde{\Omega}_i(t')-{\Omega}_i(t')\right)}\leq0.
     \end{multline*}
     which shows that the alternative policy has a lower or equal cost compared to the original one. Also, notice that the maximum number of velocity switches in the alternative policy is $2\frac{t}{d_{\min}}+4$, since there can be a maximum of $\frac{t}{d_{\min}}+1$ visits to targets with maximum speed equal to one and, for each visited target, the alternative policy has at most 2 velocity switches, plus one switch to match the initial position and another to match the terminal position of the original policy.
\end{proof}

One way to interpret this proposition is, if you look ahead the next $T$ units of time (where $T$ might represent the period of a periodic solution), there is an optimal trajectory that has its controls in the set $\{1,0,1\}$. This is the kind of trajectory that we will be pursuing in the remainder of this paper. Also, notice that even though we were not able so far to prove that there exists some optimal control solution with the same structure when the targets are not necessarily isolated, the same structure can still be used but without the guarantee of optimality.
\section{STEADY STATE PERIODIC SCHEDULES}
\label{sec:steady_state}

As stated in Sec. \ref{sec:intro}, the goal of the present work is to provide tools for analyzing the steady state behavior of the covariance matrices $\bar{\Omega}_i$. This approach contrasts with \cite{lan2014variational,cassandras2013optimal,pinto2019monitoring}, where only the transient behavior was analyzed and, therefore, the number of parameters necessary to represent the trajectory grew as the time-horizon grew. The approach here presented is particularly interesting because it captures the long-term mean estimation error while only needing to optimize the parameters that describe a single period of the trajectory.

If the agents' trajectories are constrained to be periodic, we know that $\eta_i(t)$, as defined in \eqref{eq:def_eta}, will also be periodic and, therefore, the Ricatti equation for this model, as presented in \eqref{eq:dynamics_omega}, is periodic. Before proceeding to the computation of the steady state covariance, we give a few natural assumptions on the system.
\begin{assumption}
    The pair $(A_i,H_i)$ is detectable, for every $i\in\{1,...,M\}$.
\end{assumption}
\begin{assumption}
    $Q_i$ and the initial covariance matrix $\Sigma_i(0)$ are positive definite, for every $i\in\{1,...,M\}$.
\end{assumption}

Following a procedure similar to the one used in the proof of {\it Lemma 9} in \cite{le2010scheduling}, we show that, when target $i$ is visited for at least a finite amount of time, the Riccati equation for that target \eqref{eq:dynamics_omega} converges to a unique periodic solution. Note that a solution $\bar\Omega_i$ to \eqref{eq:dynamics_omega} is said to be stabilizing if, for any solution $\Omega_i$ of \eqref{eq:dynamics_omega} with symmetric non-negative initial conditions, $\lim_{t\rightarrow \infty}\lambda_{\max}\left(\bar{\Omega}_i-\Omega_i\right)=0$, where $\lambda_{\max}(.)$ is the eigenvalue if maximum absolute value of a matrix.

\begin{proposition}
    \label{prop:unique_attractive_sol_riccati_eq}
    If $\eta_i(t) > 0$ for some interval $[a,b]\in[0,T]$ with $b>a$, then, under Assumption 1, there exists a non-negative stabilizing $T$-periodic solution to \eqref{eq:dynamics_omega}.
\end{proposition}
\begin{proof}
    According to \cite[p.~130]{bittanti1984periodic}, a pair $(A_i,\eta_i(t)H_i)$ of a periodic system is detectable if and only if for every eigenpair $(x,\lambda)$ with $x\neq0$,
    \begin{equation*}
        A_ix=\lambda x \implies \exists\ t\in[0,T]\ s.t.\ \eta_i(t)e^{\lambda t}H_ix\neq0
    \end{equation*}
    Notice that, due to Assumption 1, for any eigenvector $x$ of $A_i$, $H_ix\neq0$, therefore when $\eta_i(t)>0$ (i.e. any $t\in[a,b]$), $\eta_i(t)e^{\lambda t}H_ix\neq0$, which implies that $(A_i,\eta_i(t)H_i)$ is detectable. Therefore, the collorary to Theorem 3 in \cite[p.~95]{nicolao1992convergence} states shows that there exists a non negative $T$-periodic solution to \eqref{eq:dynamics_omega}, $\bar{\Omega}_i(t)$, and 
    \begin{equation*}
        \lim_{t\rightarrow \infty}(\Omega_i(t)-\bar{\Omega}_i(t))=0
    \end{equation*}
    for any positive definite initial condition $\Omega_i(0)$.
\end{proof}

Notice that we can always design a periodic trajectory such that every target is visited for at least a finite time interval and therefore, $\eta_i(t)>0$ for some interval. Defining $\bar{\Omega}_i(t)$ as the unique periodic solution to \eqref{eq:dynamics_omega} and $\Omega_i(t)$ as the solution for some non negative initial conditions $\Omega_i(0)$ we know that, since $\bar{\Omega}_i(t)$ is the unique stabilizing solution of \eqref{eq:dynamics_omega},
\begin{equation*}
    \forall \delta>0, \exists\ t_0\ s.t.\ \norm{\bar{\Omega}_i(t)-{\Omega}_i(t)}\leq\delta,\ \forall t\geq t_0,
\end{equation*}
which implies that
\begin{equation*}
    \lim_{t\rightarrow \infty}\frac{1}{t}\int_{t_0}^t |\tr(\tilde{\Omega}_i(t')-{\Omega}_i(t'))|\dtprime\leq\delta
\end{equation*}
and, since we know that, for a finite period of integration $t_0$, $\lim_{t\rightarrow \infty}\frac{1}{t}\int_{0}^{t_0} |\tr(\tilde{\Omega}_i(t')-{\Omega}_i(t'))|\dtprime=0$, we conclude that
\begin{equation}
    \label{eq:infinite_time_steady_state_periodic}
    \lim_{t\rightarrow \infty}\frac{1}{t}\int_{0}^t |\tr(\tilde{\Omega}_i(t')-{\Omega}_i(t'))|\dtprime\leq\delta.
\end{equation}

Equation \eqref{eq:infinite_time_steady_state_periodic} implies that, for any initial condition on the covariance matrix, if we apply a periodic schedule for the agents such that every target is visited at least once, after sufficient time, the cost given by \eqref{eq:opt_objective} will become arbitrarily close to the mean cost over time of the steady state periodic solution associated to that same periodic trajectory. This implies that if we optimize the steady state solution $\bar{\Omega}_i$, the cost of the solution starting at any arbitrary initial condition will asymptotically approach that of the steady state one as time evolves.

Consider now the motion of the agents. The result in Proposition \ref{prop:optimal_control} implies that when the targets are isolated there is an optimal control policy such that $u_j(t)\in\{-1,0,1\}$, even if the trajectory is constrained to be periodic. This property allows the optimal trajectory to be described by a finite set of parameters, as in \cite{cassandras2013optimal,pinto2019monitoring}. In this work in particular we are looking into periodic trajectories and, hence, this property implies that the movement in each period of agent $j$ consists of a sequence of dwelling at the same position for some duration of time followed by moving at maximum speed to another location. Therefore, one period of the trajectory of an agent $j$ can fully be described by the following set of parameters:
\begin{enumerate}
\item $T$, the period of the trajectory.
\item $s_{j}(0)$, the initial position.
\item $\omega_{j,p}$, $p=1,...,P_j$, the normalized dwelling times for agent $j$, i.e., the agent dwells for $\omega_{j,p}T$ units of time before it moves with maximum speed for the $p$-th time in the cycle.
\item $\tau_{j,p}$, $p=1,...,P_j$, the normalized movement times for agent $j$, i.e., the agent $j$ moves for $\tau_{j,p}T$ units of time to the right (if $p$ is odd) or to the left (if $p$ is even) after dwelling for $\omega_{j,p}T$ units of time in the same position.
\end{enumerate}

The following five constraints are enforced to ensure periodicity of the trajectory and consistency of its parameters. The last two constraints ensure, respectively, that the total time that the agents spend moving will be less or equal than a period and that the amount that over the course of a period, each agent will return to its initial position.
\begin{equation}
\label{eq:parameters_constraints}
\begin{gathered}
    \tau_{j,m}\geq 0,\ \omega_{j,m}\geq 0,\ T\geq 0,\  \sum_{m=1}^{P_j}(\tau_{j,m}+\omega_{j,m})\leq 1  \\ \sum_{m=1}^{P_j}(-1)^m \tau_{j,m}=0.
\end{gathered}    
\end{equation}

Notice that this description does not exclude transitions of $u_j$ of the kind $\pm1\rightarrow \mp 1$ and $\pm1\rightarrow 0 \rightarrow \pm1$, since it allows $\omega_{j,m}=0$ and $\tau_{j,m}=0$. This parameterization defines a hybrid system in which the dynamics of the agents remain unchanged between events and abruptly switch when an event occurs. Events are given by a change in control value at completion of movement and dwell times. Note that these may occur simultaneously, for instance, if the dwell time is zero (representing a switch of control from $\pm 1$ to $\mp 1$). Given this parameterization, we use an approach analogous to \cite{cassandras2013optimal,pinto2019monitoring} in which IPA is used to calculate the stochastic gradient estimate of the cost function with respect to the parameters defining the trajectories and then the gradient is used in a gradient descent scheme to optimize the cost function. 

\section{OPTIMIZATION OF THE PERIODIC TRAJECTORY}
\label{sec:hybrid_sys}
In the previous section, we described how, in an environment with isolated targets, the optimal steady state trajectory computation can be framed as an optimization of a finite set of parameters that represent the trajectory. We also showed that if this trajectory is used from any arbitrary initial condition, the mean cost will become arbitrarily close to the steady state cost. In this section we take advantage of the fact that, if every target is observed at least once, the Riccati equation is globally attractive in order to compute the derivative of the steady state solution of the Riccati equation with respect to all the parameters that are part of it. These can be used in a gradient descent scheme to obtain locally optimal steady state solution. Note that the suboptimality of the gradient descent in the context of persistent monitoring, along with an approach to converge to better local optima, is discussed in \cite{zhou2018optimal}. 

In this work, we take advantage of Infinitesimal Perturbation Analysis (IPA) to compute these gradients.
IPA is a tool for estimating stochastic gradients of hybrid system states and event times with respect to given system parameters. These estimates, under mild assumptions on the distribution of the random processes involved, have the interesting property of being unbiased and distribution invariant \cite{cassandras2010perturbation}. IPA is particularly attractive due to its event driven nature, i.e., the equations used in the computation of the parameters only need to be updated when some event (e.g. a transition of the discrete mode of the system) happens, which means that effort for updating the equations scales linearly with the number of events (rather than exponentially with the number of targets and agents).
\subsection{IPA Formulation}
By defining $q=t/T$, \eqref{eq:dynamics_omega} can be rescaled as
\begin{multline}
\label{eq:scaled_riccati_diff_eq}
\dot{\Omega}_i(q)=\frac{d{\Omega}_i(q)}{dq} = T(A\Omega_i(q)+\Omega_i(q)A^T\\+Q-\eta_i(q)\Omega_i(q)G\Omega_i(q)),
\end{multline}

In order to optimize the parameters of the agent trajectories using gradient descent, we need the gradient of the cost with respect to these parameters. Taking the derivative of \eqref{eq:opt_objective}, we have that for any parameter $\theta$
\begin{equation}
    \frac{\partial J}{\partial \theta} = \sum_{i=1}^{M}\int_{0}^{1}\frac{\partial{\text{tr}(\Omega_i(q))}}{\partial \theta} dq.
\end{equation}

Using IPA, we derive the ordinary differential equations for which the desired gradient $\frac{\partial{\Omega_i(t)}}{\partial \theta}$ is a solution. Note that in this paper we sidestep the issue of whether or not these gradients exist.
We know that there are sets of parameters for which the gradient does not exist (imagine, for instance, a set of parameters for which one of the targets is never visited and the dynamics of this target are unstable, therefore, its covariance diverges as time goes to infinity). 
However, experience and prior results support the assumption that these gradients do indeed exist in the interior of the set of parameters for which each target is visited at least once. 

Computing the derivative of $\Omega_i$ with respect to any parameter $\theta$ (except for $T$) yields
\begin{multline}
    \label{eq:derivative_omega_i}
    \frac{\partial \dot{\bar\Omega}_i(q)}{\partial \theta}-T\biggl(A\frac{\partial\bar{\Omega}_i(q)}{\partial \theta}+\frac{\partial\bar{\Omega}_i(q)}{\partial \theta}A^T\\-\eta_i(q)\bar{\Omega}_i(q)G\frac{\partial\bar{\Omega}_i(q)}{\partial \theta}-\eta_i(q)\frac{\partial\bar{\Omega}_i(q)}{\partial \theta}G\bar{\Omega}_i(q)\biggr) =\\ T\frac{\partial \eta_i(q)}{\partial \theta}\bar{\Omega}_i(q)G\bar{\Omega}_i(q),
\end{multline}
where one should look at $\frac{\partial {\bar\Omega}_i}{\partial \theta}$ as the unknown function which we are trying to solve for. In this expression, the term $\eta_i(q)$ is fully determined by the agent's trajectory parameters, the steady state covariance matrix $\bar{\Omega}_i(q)$ as described in the previous section and explicit expressions for the term $\frac{\partial \bar{\Omega}_i}{\partial \theta}$ will be given in the next subsection. 

Since \eqref{eq:derivative_omega_i} does not fully determine a unique solution (different initial conditions $\frac{\partial \bar{\Omega}_i}{\partial \theta}(0)$  will yield different solutions), we need extra conditions to determine the partial derivatives of the covariance matrix. Since $\bar{\Omega}_i(q)$ is periodic, $\frac{\partial \bar{\Omega}_i}{\partial \theta}$ must also be periodic. This property will allow us to uniquely determine the initial conditions for computing the derivative $\frac{\partial \bar{\Omega}_i}{\partial \theta}$, as discussed in the following.

Define the problem:
\begin{equation}
    \label{eq:homogeneous_derivative_lyapunov_eq}
    \dot{\Sigma}_H(q)-T\left(A-\eta_i(q)\bar{\Omega}_i(q)G\right)\Sigma_H(q)=0,\ \Sigma_H(0)=I
\end{equation}
and let $\Sigma_{ZI}$ be the solution of \eqref{eq:derivative_omega_i} with the zero matrix as the initial conditions. Also, let $\Sigma_H$ denote the solution of the homogeneous version of \eqref{eq:derivative_omega_i} with the identity matrix as the initial condition. Then, the initial conditions matrix $\Lambda$ that yields a periodic solution of  \eqref{eq:derivative_omega_i} is such that \cite{reid1946matrix}:
\begin{equation}
    \label{eq:lyapunov_equation}
    \Lambda=\Sigma_H(1)\Lambda\Sigma_H^T(1)+\Sigma_{ZI}(1),
\end{equation}
which has at least one solution $\Lambda$ if $\frac{\partial \bar{\Omega}_i}{\partial \theta}$ exists. The following proposition states sufficient conditions for uniqueness.

\begin{proposition}
    \label{prop:unique_sol_lyapunov_eq}
    Under the following conditions:
    \begin{itemize}
        \item $\Sigma_H$ is a solution of \eqref{eq:homogeneous_derivative_lyapunov_eq};
        \item Assumptions 1 and 2 hold;
        \item target $i$ is observed at least once in the period $T$;
        \item there exists a solution to \eqref{eq:lyapunov_equation};
    \end{itemize}
    Then, the solution to \eqref{eq:lyapunov_equation} is unique.
\end{proposition}
\begin{proof}
If $\Lambda$ and $\tilde{\Lambda}$ are solutions of \eqref{eq:lyapunov_equation}, then
\begin{equation}
    \Lambda-\tilde{\Lambda} = \Sigma_H(1)\left(\Lambda-\tilde{\Lambda}\right)\Sigma_H^T(1) 
\end{equation}
which is equivalent to
\begin{equation}
    \label{eq:vectorization_difference_solutions_Lyapunov}
    \Vec{\left(\Lambda-\tilde{\Lambda}\right)} = \left(\Sigma_H(1)\otimes \Sigma_H(1)\right) \Vec{\left(\Lambda-\tilde{\Lambda}\right)}
\end{equation}
Notice that $\Lambda=\tilde{\Lambda}$ is a solution of \eqref{eq:vectorization_difference_solutions_Lyapunov} and it is the unique solution if and only if $1$ is not an eigenvalue of $\Sigma_H(1)\otimes \Sigma_H(1)$, and its eigenvalues are in the form $\mu_1\mu_2$, where $\mu_1$ and $\mu_2$ are distinct eigenvalues of $\Sigma_H(1)$ \cite{zhang2011matrix}.

In the following we show that all the eigenvalues of $\Sigma_H(1)$ have absolute value lower than one. For that, first notice that since Q is positive definite, then $\bar{\Omega}_i$ must be positive definite and hence, invertible. We can define
\begin{equation*}
    \mathcal{I} = \bar{\Omega}_i^{-1},
\end{equation*}
and, since $\dot{\mathcal{I}}=-\Omega_i^{-1}\dot\Omega_i\Omega^{-1}_i=-\mathcal{I}\dot\Omega_i\mathcal{I}$, the dynamics of $\mathcal{I}$ can be expressed as:
\begin{equation}
    \dot{\mathcal{I}} = -T(\mathcal{I} A + A^T \mathcal{I} + \mathcal{I} Q \mathcal{I}-\eta_i G).
\end{equation}

Therefore, if we define the Lyapunov Function $V = \Sigma_H^T\mathcal{I}\Sigma_H$, we know that:
\begin{equation}
\begin{aligned}
    \frac{d}{dq}\left(\Sigma_H^T\mathcal{I}\Sigma_H\right)&=\Sigma_H^T\left(\mathcal{I}A+AˆT\mathcal{I}+2\eta_i G+\dot{\mathcal{I}}\right)\Sigma_H\\
    &=-\Sigma_H^T\mathcal{I}Q\mathcal{I}\Sigma_H.
\end{aligned}
\end{equation}

By integrating the previous relation and using the fact that $\mathcal{I}$ is periodic with period one, $\Sigma_H(0)=I$ and assumption 2, we know that
\begin{multline}
    \Sigma_H(1)\mathcal{I}(0)\Sigma_H(1)-\mathcal{I}(0)=\\-\int_0^1 \Phi(q,0)^T\mathcal{I}Q\mathcal{I}\Phi(q,0) \dq < 0,
\end{multline}
where $\Phi(q_1,q_2)$ is the transition matrix of the system \eqref{eq:homogeneous_derivative_lyapunov_eq} betwen times $q_1$ and $q_2$.  
Since $\mathcal{I}$ is positive definite, we know that the system is stable and therefore the absolute value of all the eigenvalues of $\Sigma_H(1)$ are lower than 1. This implies that $\mu_1\mu_2<1$ and therefore the solution of \eqref{eq:lyapunov_equation} has to be unique.

\end{proof}

The Lyapunov Equation in \eqref{eq:lyapunov_equation} can be efficiently solved for low-dimensional systems using the algorithm proposed in \cite{barraud1977numerical} and implemented in MATLAB function $dlyap$. The needed derivative can then be computed as:
\begin{equation}
    \frac{\partial {\bar\Omega}_i(q)}{\partial \theta} = \Sigma_H^T(q)\Lambda\Sigma_H(q)+\Sigma_{ZI}(q).
\end{equation}

The same discussion also holds for the parameter $T$, with the differential equation \eqref{eq:derivative_omega_i} replaced by:
\begin{multline}
    \label{eq:derivative_T}
    \frac{\partial \dot{\bar\Omega}_i(q)}{\partial T}-T\biggl(A\frac{\partial\bar{\Omega}_i(q)}{\partial T}+\frac{\partial\bar{\Omega}_i(q)}{\partial T}A^T\\-\eta_i(q)\bar{\Omega}_i(q)G\frac{\partial\bar{\Omega}_i(q)}{\partial T}-\eta_i(q)\frac{\partial\bar{\Omega}_i(q)}{\partial T}G\bar{\Omega}_i(q)\biggr) =\\ A\bar{\Omega}_i(q)+\bar{\Omega}_i(q)A^T+Q-\left(\eta_i(q)+T\frac{\partial \eta_i(q)}{\partial T}\right)\bar{\Omega}_i(q)G\bar{\Omega}_i(q),
\end{multline}
which is associated to the same homogeneous equation \eqref{eq:homogeneous_derivative_lyapunov_eq}.

\subsection{Computation of $\frac{\partial \eta_i(q)}{\partial \theta}$}
Looking back to \eqref{eq:derivative_omega_i}, in order to give a complete procedure for computing the derivative $\frac{\partial J}{\partial \theta}$ when it exists, the only component left is to compute the derivative $\frac{\partial \eta_i(q)}{\partial \theta}$. Using \eqref{eq:def_eta}, we know that 
\begin{equation}
    \frac{\partial \eta_i(q)}{\partial \tau_{j,m}} = -\frac{I_j(s_j-x_i)}{r_j}\frac{\partial{s_j(q)}}{\partial \tau_{j,m}}, 
\end{equation}
where
\begin{equation}
    I_j(\alpha)=
    \begin{cases}
        +1, & 0<\alpha<r_j, \\
        -1, & -r_j<\alpha<0, \\
        0, & |\alpha|>r_j.
    \end{cases}
\end{equation}

As a side note, since $\gamma_{i,j}$ is not differentiable at $\alpha=0$, we can use the concept of subgradient and use any value between $-1$ and $1$ for $I_j(0)$. Similarly, 

\begin{equation}
    \frac{\partial \eta_i(q)}{\partial \omega_{j,m}} = -\frac{I_j(s_j-x_i)}{r_j}\frac{\partial{s_j(q)}}{\partial \omega_{j,m}} ,
\end{equation}
\begin{equation}
    \frac{\partial \eta_i(q)}{\partial s_{j}(0)} = -\frac{I_j(s_j-x_i)}{r_j}\frac{\partial{s_j(q)}}{\partial s_{j}(0)} ,
\end{equation}
\begin{equation}
    \frac{\partial \eta_i(q)}{\partial T} = -\sum_{j=1}^N\frac{I_j(s_j-x_i)}{r_j}\frac{\partial{s_j(q)}}{\partial T} .
\end{equation}

In order to compute $\frac{\partial{s_j(q)}}{\partial \theta}$ for some parameter $\theta$ we will explicitly write the position $s_j(q)$ as a function of this parameter. As already discussed, IPA is event-driven in nature. These events, for our parameterization, are the instants when the trajectory presents a change in the velocity, at the end of dwell times or movement times. The dynamics of the derivatives may experience discontinuities at these specific event times.  The order of the events is defined in such a way that initially the agent dwells, then it moves right, then dwells again, followed by moving left and repeat this sequence until the number of events reaches $2P_j$, where $P_j$ is a designer-defined parameter that indicates the maximum number of direction switches the agent $j$ can experience in its trajectory. Note that the value of $P_j$ is upper bounded when the targets are isolated and Proposition \ref{prop:optimal_control} gives an upper bound for $P_j$ as a function of the period $T$. Also, notice that under this definition events are agent-specific and can happen at different times for different agents.

The position of agent $j$ at normalized time $q$, after the $k$-th event and before the $k+1$-th is
\begin{equation}
s_j(q) -s_j(0)=
\begin{cases}
    T\biggl((-1)^{k/2+1}\biggl(q-\sum_{p=1}^{k/2-1}(\tau_{j,p}+\omega_{j,p})\\ {\ \ }+\omega_{j,\frac{k}{2}}\biggr)+\sum_{p=1}^{k/2}(-1)^{p+1}\tau_p\ \biggl),\ k\ \text{even},\\
    T\sum_{p=1}^{\frac{k-1}{2}}(-1)^{p+1}\tau_{j,p},\ k\ \text{odd}.
\end{cases}
\end{equation}

Therefore,
\begin{equation}
    \frac{\partial s_j}{\partial \tau_{j,m}} = 
 \begin{cases}
    \left((-1)^{\frac{k}{2}+1}+(-1)^p\right)T,&\ m<\frac{k}{2},\ k\ \text{even},\\
    (-1)^{m+1}T,\ m\leq\frac{k-1}{2},&\ k\ \text{odd},
 \end{cases}
\end{equation}

\begin{equation}
    \frac{\partial s_j}{\partial \omega_{j,m}} =
    \begin{cases}
        1,\ m<\frac{k}{2},&\ k\ \text{even},\\
        0&,\ \text{otherwise},
    \end{cases}
\end{equation}
    
\begin{equation}
    \frac{\partial s_j(q)}{\partial T} = \frac{s_j(q)-s_j(0)}{T},
\end{equation}

\begin{equation}
    \frac{\partial s_j}{\partial s_j(0)}=1.
\end{equation}

\subsection{Complete Optimization Procedure}
In this subsection, we summarize our approach by gathering all the components of the optimization procedure into a single algorithm. First, we define the parameter set $\Theta_j^l$:
\begin{equation}
    \Theta_j^l = [s_{j}(0),\tau_{j,1}^l,...,\tau_{j,P_j}^l,\omega_{j,1}^l,...,\omega_{j,P_j}^l],    
\end{equation}
where the upper index $l$ refers to the step number in the gradient descent, i.e.,
\begin{equation}
    {\Theta}_{j}^{l+1}= \text{proj}\left(\Theta_{j}^{l}-\kappa_l\frac{\partial J}{\partial{{\Theta}}_{j}}\right),
\end{equation}
where {\it{proj}} represents the projection into the convex set defined by the constraints in \eqref{eq:parameters_constraints} and $\kappa_l$ is the gradient descent step size. Algorithm \ref{alg:agents_optimization} describes the complete optimization procedure.
\begin{algorithm}
\caption{Agents' Trajectory Optimization}
\label{alg:agents_optimization}
\begin{algorithmic}[1]
\Procedure{Gradient Descent}{}
\State {\bf Input:} $\Theta_1^0$,..,$\Theta_N^0,T^0$, 
\State $||\nabla J|| \leftarrow \infty$
\State $l\leftarrow 0$
\While {$||\nabla J||>\epsilon$}
\State $\left[\frac{\partial J}{\partial{\Theta}_{1}},...,\frac{\partial J}{\partial{\Theta}_{N}},\frac{\partial J}{\partial T}\right]\leftarrow$IPA($\Theta_1^l,...,\Theta_N^l,T^l$)
\For {$j$ from $1$ to $N$}
\State ${\Theta}_{j}^{l+1}\leftarrow \text{proj}\left(\Theta_{j}^{l}-\kappa_l\frac{\partial J}{\partial{{\Theta}}_{j}}\right)$
\EndFor
\State $T^{l+1}\leftarrow T^l-\kappa_l\frac{\partial J}{\partial T}$
\State $||\nabla J|| \leftarrow \frac{1}{\kappa_k}\sqrt{\left(\frac{\partial J}{\partial T}\right)^2+\sum_{j=1}^N\norm{\Theta_k^{l+1}-\Theta_k^{l}}^2}$
\State $l\leftarrow l+1$

\EndWhile
\State {\bf Output:} $\underline{\theta}_1^l,...,\underline{\theta}_N^l,\underline{\omega}^l_1,...,\underline{\omega}^l_N$
\EndProcedure
\State
\Procedure {IPA}{}
\State{\bf Input}: $\Theta_1,...,\Theta_N,T$
\For{$j$ from $1$ to $N$}
    \State$\frac{\partial J}{\partial \Theta_j}\leftarrow 0$
\EndFor
\State$\frac{\partial J}{\partial T}\leftarrow 0$
\State Compute $s_1(q),...,s_N(q)$ from the parameterization
\For{$i$ from $1$ to $M$}
\State Compute $\bar{\Omega}_i(q)$ by running to \eqref{eq:scaled_riccati_diff_eq} until it converges to a periodic solution
\State Compute $\Sigma_H^i$ from \eqref{eq:homogeneous_derivative_lyapunov_eq}
\For{$j$ from $1$ to $N$}
\For{every $\theta$ in $\Theta$}
\State Solve \eqref{eq:derivative_omega_i} with zero initial conditions to compute $\Sigma^{\theta,i}_{ZI}$
\State Compute $\Lambda^{\theta,i}$ using \eqref{eq:lyapunov_equation}
\State $I \leftarrow \int_0^1\tr\left((\Sigma_H^i)^T\Lambda^{\theta,i}\Sigma_Hî+\Sigma_{ZI}^{\theta,i}\right)dq$
\State $\frac{\partial J}{\partial \theta_j}\leftarrow \frac{\partial J}{\partial\theta_j}+I$
\EndFor
\EndFor
\State Solve \eqref{eq:derivative_T} with zero initial conditions to compute $\Sigma^{T,i}_{ZI}$
\State Compute $\Lambda^{T,i}$ using \eqref{eq:lyapunov_equation}
\State $I \leftarrow \int_0^1\tr\left((\Sigma_H^i)^T\Lambda^{T,i}\Sigma_H^i+\Sigma_{ZI}^{T,i}\right)dq$
\State $\frac{\partial J}{\partial T}\leftarrow \frac{\partial J}{\partial T}+I$

\EndFor
\State {\bf Output:} $\frac{\partial J}{\partial{\Theta}_{1}},...,\frac{\partial J}{\partial {\Theta}_{j}},\frac{\partial J}{\partial T}$
\EndProcedure
\end{algorithmic}
\end{algorithm}

In this paper, a procedure for obtaining $\Theta_j^0$ is not discussed. One essential condition for this initial configuration is that every target is visited at least once for a finite amount of time, as discussed in Sec. \ref{sec:steady_state}, otherwise the covariance matrices will not converge to a steady-state solution. Although providing efficient initial parameters for the optimization is a topic that we are still investigating, one possible way to address it would be to use the transient analysis given in \cite{pinto2019monitoring}, possibly augmenting it with the technique 
proposed in \cite{zhou2018optimal}, 
where the cost function was augmented to provide a larger exploration of the environment by the gradient descent algorithm.

Also, it is worth noting that although we have already shown that the steady state solution of the periodic Riccati equation is globally attractive, no convergence rate was indicated. There are, however, various alternative numerical methods that can provide guaranteed convergence that could replace line 21 in Alg. \ref{alg:agents_optimization}. We refer the reader to \cite{varga2013computational} for a more complete discussion of these methods.

\section{Simulation Results}
\label{sec:results}
In this section, we demonstrate the results of Algorithm \ref{alg:agents_optimization} in a two different scenarios, one with only one agent and two targets and a second one with five targets and two agents. All targets $i$ have the same state dynamics evolving according to \eqref{eq:dynamics_phi} with
\begin{align*}
    A_i = \begin{bmatrix} -1 & -0.1 \\ -0.1 & 0.01 \end{bmatrix}, \quad Q_i = \textrm{diag}(1,1),
\end{align*}
and the same observation model as in \eqref{eq:observation_model_ij} with
\begin{align*}
    H_i = \textrm{diag}(1,1), \quad R_i = \textrm{diag}(1,1),\quad r_j=0.9.
\end{align*}

A constant  descent stepsize was used ($\kappa_l = \kappa_0$) and, order to provide an index convention for the targets, we define that $x_1<x_2<...<x_M$.
\subsection{One agent, two targets}
In the first scenario, with one agent and two targets, the following set of initial conditions was used:
\begin{align*}
    s_1^0(0)=0,\quad T^0=6,\quad \tau_1^0=[0.2,0.4,0.2],
\end{align*}
\begin{align*}
    \omega_1^0=[0.05,0.05,0.05],
\end{align*}
and the gradient descent step size was set to $\kappa_0=0.02$.

\begin{figure*}[htp!]
    \centering
    \begin{subfigure}{0.32\textwidth}
        \centering\includegraphics[width=\textwidth]{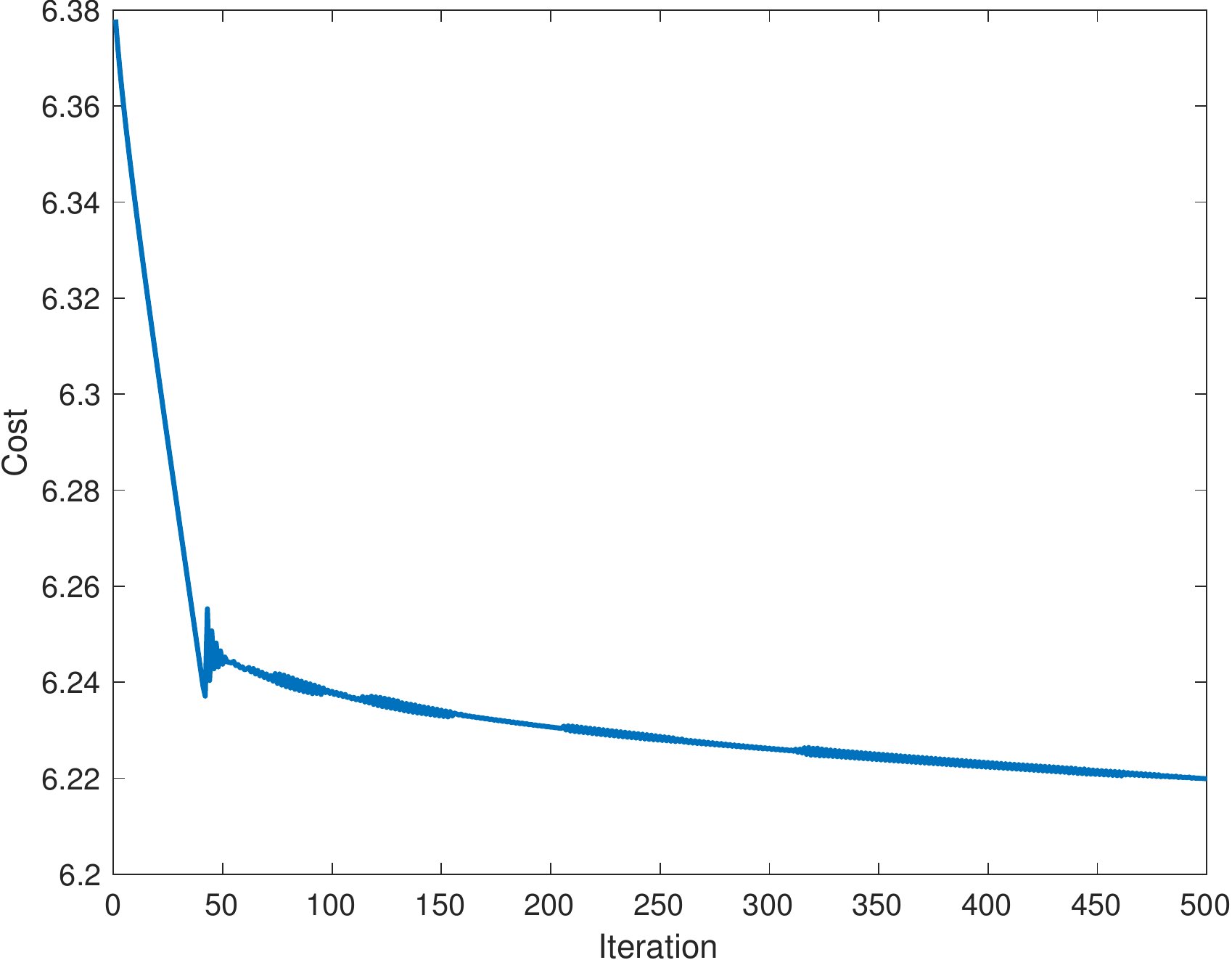}
        \caption{Cost vs. iteration number}
        \label{fig:cost}
    \end{subfigure}
    \begin{subfigure}{0.32\textwidth}
        \centering\includegraphics[width=\textwidth]{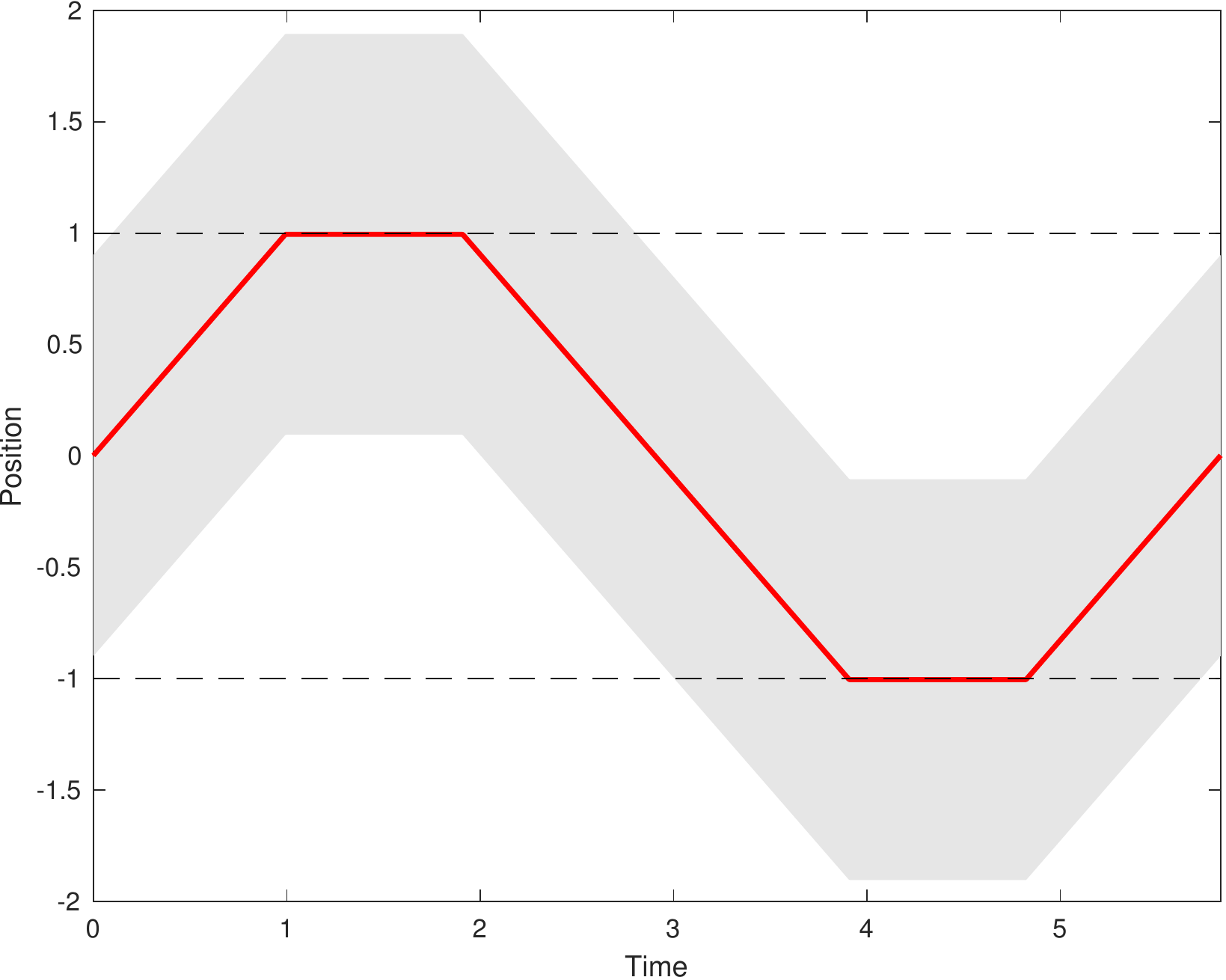}
        \caption{Agent trajectories at final iteration}
        \label{fig:position_1_target}
    \end{subfigure}
    \begin{subfigure}{0.32\textwidth}
        \centering\includegraphics[width=\textwidth]{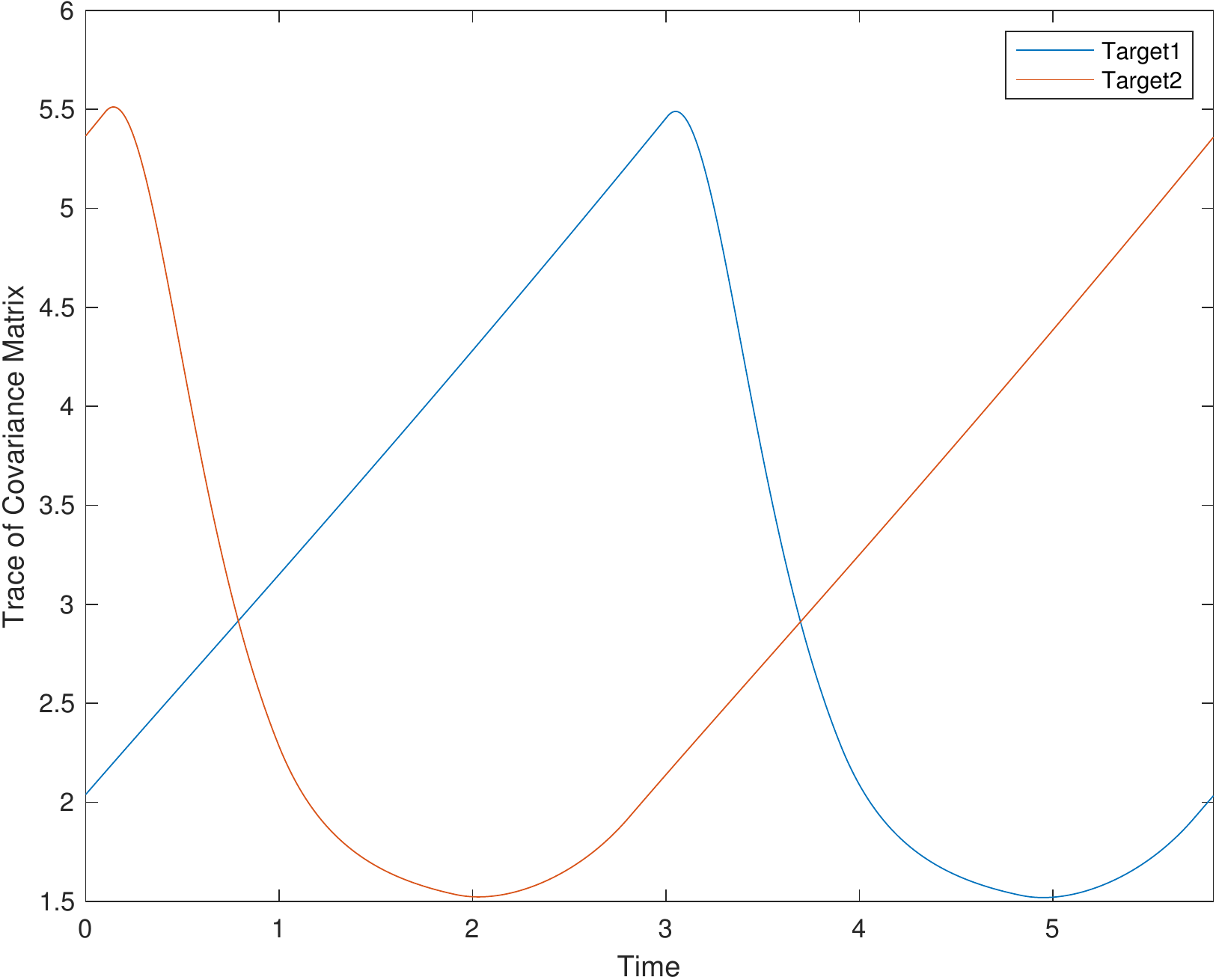}
        \caption{Trace of the covariance for each target}
        \label{fig:covariance_1_target}
    \end{subfigure}
    \caption{Results of a simulation with one agent and two targets. (a) Evolution of the overall cost as a function of iteration number on the gradient descent. (b) Trajectories of the agent at the final iteration. The dashed lines indicate the positions of the targets and the grey shaded area the visibility region of the agent. (c) Evolution of the trace of the estimation covariance matrices of the two targets. }
    \label{fig:results_1_target}
\end{figure*}

Figure \ref{fig:results_1_target} illustrates the results of running the simulation in this scenario. By analyzing Figures \ref{fig:position_1_target} and \ref{fig:covariance_1_target}, the optimized policy is such that the agent moves between the two targets and dwells on top of them in a symmetric way. The steady state covariances behave with very similar curves, but shifted in time. Also notice that the period is lower than the original one.

\subsection{Two agents and Five targets}
In this second scenario, the targets were placed in positions $x_i=1+2i,\ i=1,..,5.$ The initial parameters were the following:
\begin{align*}
    s_1^0(0)=2.7,\quad s_2(0)=6.8, \quad T^0=6, 
\end{align*}
\begin{align*}
    \tau_1^0=\tau_2^0=0.1[1,0.1,1,1,0.1,1,0.1,1,1,0.1,1],
\end{align*}
\begin{align*}
    \omega_1^0=\omega_2^0=0.0125[1,1,1,1,1,1,1,1,1,1,1],
\end{align*}
and the gradient descent step size was set to be constant, $\kappa_0=\kappa_l=0.02$.
\begin{figure*}[htp!]
    \centering
    \begin{subfigure}{0.32\textwidth}
        \centering\includegraphics[width=\textwidth]{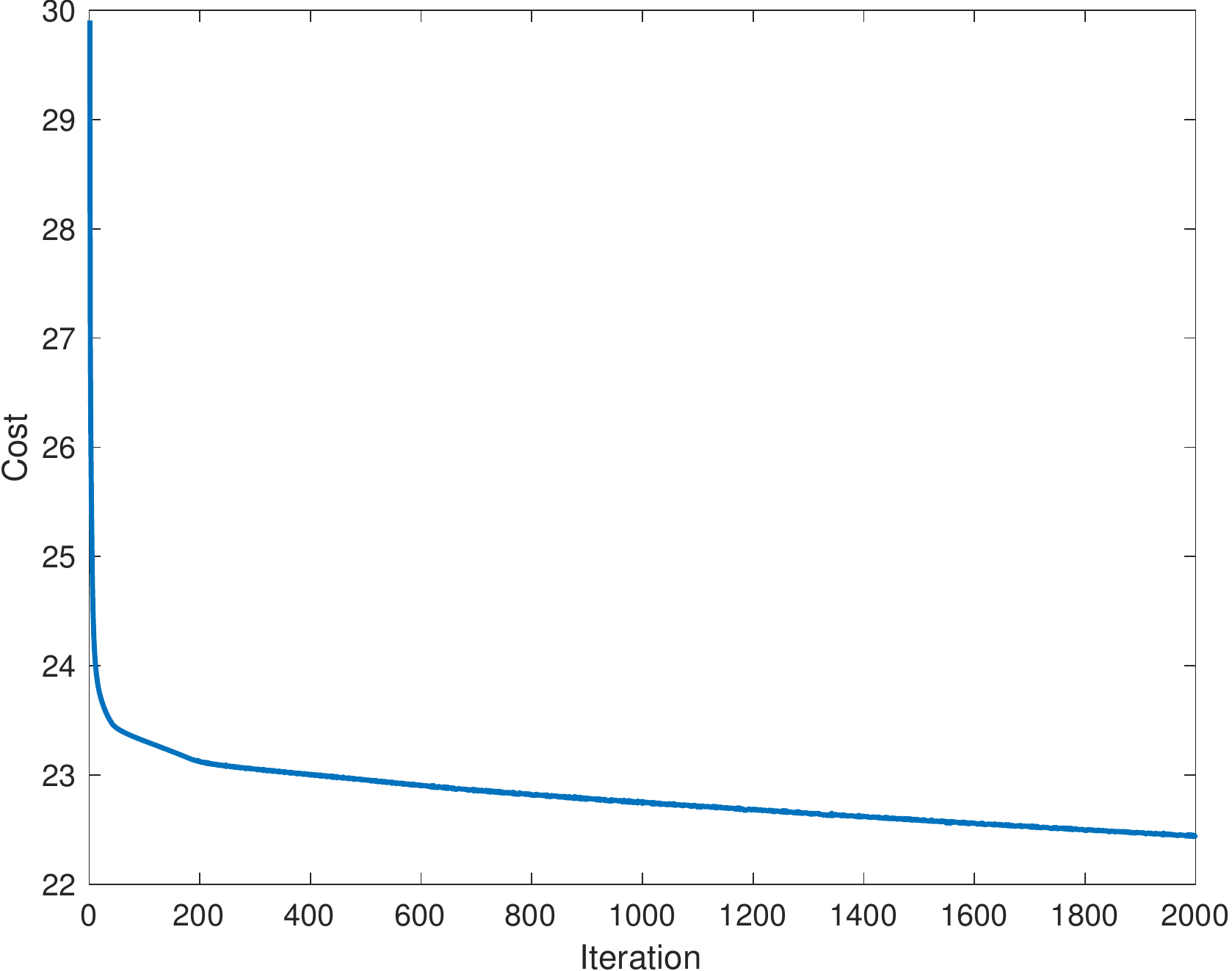}
        \caption{Cost vs. iteration number}
        \label{fig:cost_5_target}
    \end{subfigure}
    \begin{subfigure}{0.32\textwidth}
        \centering\includegraphics[width=\textwidth]{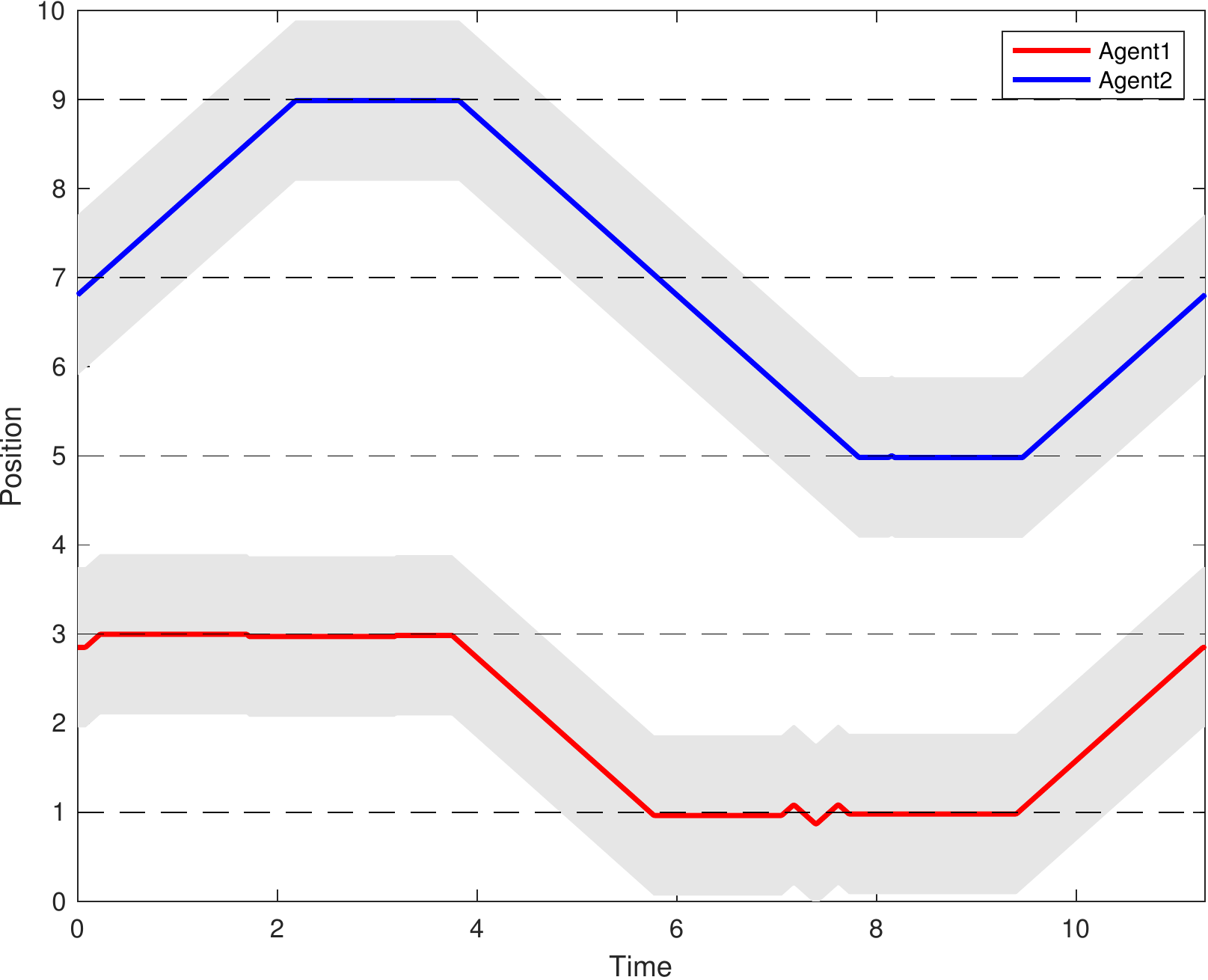}
        \caption{Agent trajectories at final iteration}
        \label{fig:position_5_target}
    \end{subfigure}
    \begin{subfigure}{0.32\textwidth}
        \centering\includegraphics[width=\textwidth]{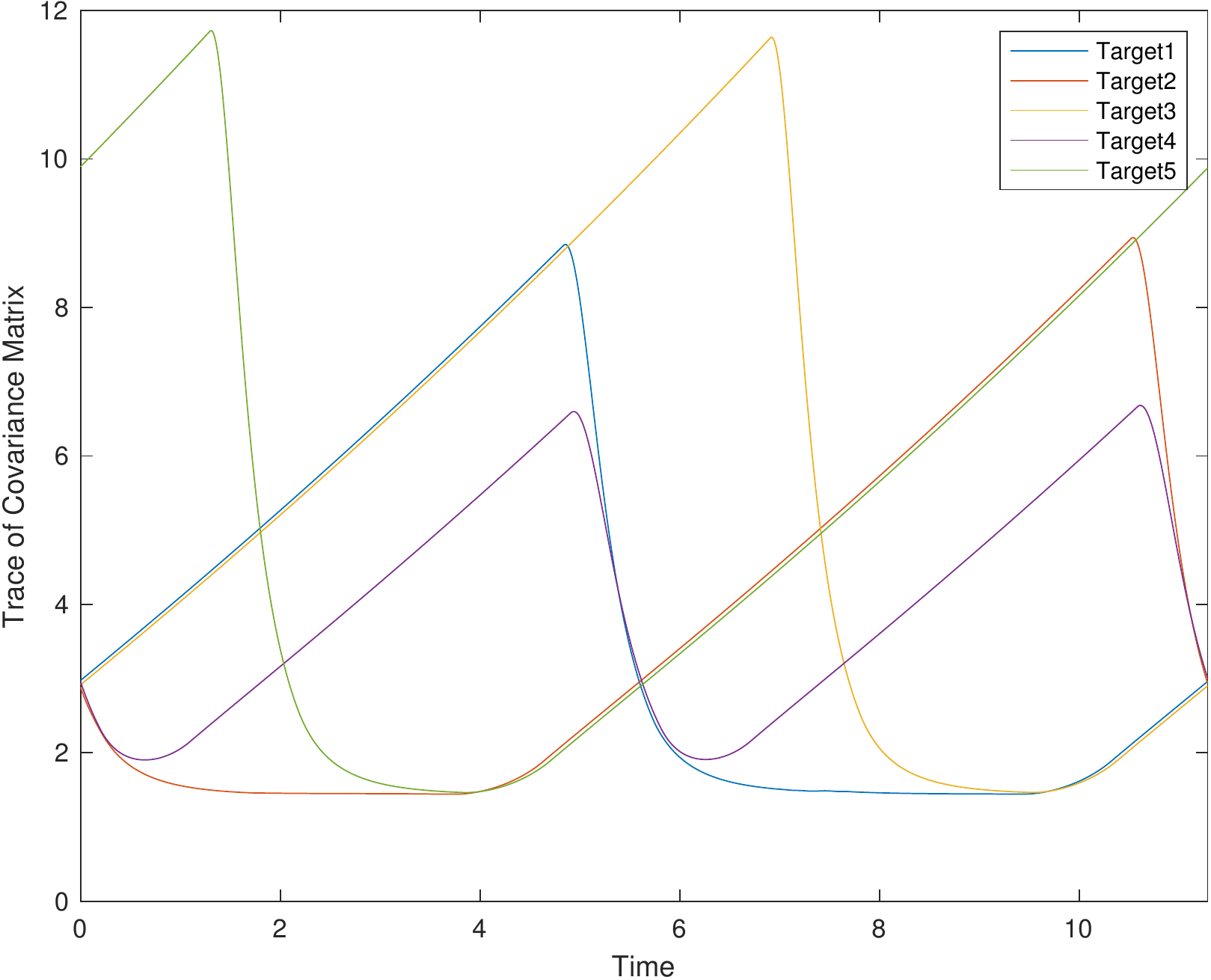}
        \caption{Trace of the covariance for each target}
        \label{fig:covariance_5_target}
    \end{subfigure}
    \caption{Results of a simulation with two agents and five targets. (a) Evolution of the overall cost as a function of iteration number on the gradient descent. (b) Trajectories of the agents at the final iteration. The dashed lines indicate the positions of the targets and the grey shaded area the visibility region of the agent. (c) Evolution of the trace of the estimation covariance matrices of the five targets. }
    \label{fig:results_5_target}
\end{figure*}

Figure \ref{fig:results_5_target} shows the results of the optimization in this scenario. Notice that even though both agents and all the targets have the same dynamical models, the solution at the last iteration of the optimization was such that one of the agents visits three of the targets and the other two of them. One interesting aspect of the trajectories of the targets in Fig. \ref{fig:position_5_target} is that, while between times 6 and 8 agent 1 makes a movement with small amplitude around target 1, the effects of this oscillatory movement are hard to notice in the trace of the covariance of target 1 in Fig. \ref{fig:covariance_5_target}. Therefore, even though it is intuitively clear that staying still rather than moving with this oscillatory behavior will lead to a lower cost solution, the difference in terms of cost is minor. Also, notice that the solution has not yet fully converged, as can be seen in Fig. \ref{fig:cost_5_target}, the results are shown this way to highlight interesting aspects of the process. The effect of the gradient descent step size (or, more generally, the descent algorithm applied) and its effect on the convergence rate, are topics of future research.

Finally, note that while the maximum number of switches in a direction allowed to each agent was set to 11, the final solution appears to have fewer because some of the movement and dwelling times in the final solution are essentially zero.
\section{Conclusions and Future Work}
\label{sec:conclusion}
In this paper, we developed a technique both to analyze and to optimize the steady state mean estimation error of a finite set of targets being monitored by a finite set of moving agents. The structure of the optimal solution allowed us to represent it in a parametric way and we provided numerical tools to optimize it in a scalable manner. Some simulation examples were provided in order to demonstrate the proposed technique. Among the open questions we plan to address in future work are the following.
\begin{itemize}
    \item Do the gradients of $\Omega_i$ with respect to the parameters that define the trajectory always exist in the interior of the set where they lead to a convergent $\Omega_i$?
    \item How can we efficiently generate initial trajectories in order to converge to global optimal points or, at least, good local optima?
\end{itemize}

Also, the simulated results highlight the interesting feature that the locally optimal solution split the set of targets into indepedent sets. That is, no targets were shared by agents. Even though this might not always hold in general, this feature motivates the future investigation of policies where only one agent observes each target and the agents would not necessarily be constrained to the same movement period. We also plan to extend the results here presented to scenarios where the agents are not constrained to a single dimensions, possibly using suboptimal parameterizations for the trajectory, as in \cite{lin2014optimal}.

\bibliographystyle{IEEEtran}
\bibliography{references.bib}

\begin{thebibliography}{10}
\providecommand{\url}[1]{#1}
\csname url@samestyle\endcsname
\providecommand{\newblock}{\relax}
\providecommand{\bibinfo}[2]{#2}
\providecommand{\BIBentrySTDinterwordspacing}{\spaceskip=0pt\relax}
\providecommand{\BIBentryALTinterwordstretchfactor}{4}
\providecommand{\BIBentryALTinterwordspacing}{\spaceskip=\fontdimen2\font plus
\BIBentryALTinterwordstretchfactor\fontdimen3\font minus
  \fontdimen4\font\relax}
\providecommand{\BIBforeignlanguage}[2]{{%
\expandafter\ifx\csname l@#1\endcsname\relax
\typeout{** WARNING: IEEEtran.bst: No hyphenation pattern has been}%
\typeout{** loaded for the language `#1'. Using the pattern for}%
\typeout{** the default language instead.}%
\else
\language=\csname l@#1\endcsname
\fi
#2}}
\providecommand{\BIBdecl}{\relax}
\BIBdecl

\bibitem{Stump:2011gv}
E.~Stump and N.~Michael, ``{Multi-robot Persistent Surveillance Planning as a
  Vehicle Routing Problem},'' in \emph{IEEE International Conference on
  Automation Science and Engineering}.\hskip 1em plus 0.5em minus 0.4em\relax
  IEEE, 2011, pp. 569--575.

\bibitem{Yu:2016fn}
J.~Yu, M.~Schwager, and D.~Rus, ``{Correlated Orienteering Problem and its
  Application to Persistent Monitoring Tasks},'' \emph{IEEE Transactions on
  Robotics}, vol.~32, no.~5, pp. 1106--1118, 2016.

\bibitem{Yu:2017iw}
X.~Yu, S.~B. Andersson, N.~Zhou, and C.~G. Cassandras, ``{Optimal Dwell Times
  for Persistent Monitoring of a Finite set of Targets},'' in \emph{American
  Control Conference}.\hskip 1em plus 0.5em minus 0.4em\relax IEEE, 2017, pp.
  5544--5549.

\bibitem{Yu:2018cf}
------, ``{Optimal Visiting Schedule Search for Persistent Monitoring of a
  Finite Set of Targets},'' in \emph{American Control Conference (ACC)}.\hskip
  1em plus 0.5em minus 0.4em\relax IEEE, 2018, pp. 4032--4037.

\bibitem{cassandras2013optimal}
C.~G. Cassandras, X.~Lin, and X.~Ding, ``{An Optimal Control Approach to the
  Multi-agent Persistent Monitoring Problem},'' \emph{IEEE Transactions on
  Automatic Control}, vol.~58, no.~4, pp. 947--961, 2013.

\bibitem{zhou2018optimal}
N.~Zhou, X.~Yu, S.~B. Andersson, and C.~G. Cassandras, ``{Optimal Event-Driven
  Multiagent Persistent Monitoring of a Finite Set of Data Sources},''
  \emph{IEEE Transactions on Automatic Control}, vol.~63, no.~12, pp.
  4204--4217, 2018.

\bibitem{lan2013planning}
X.~Lan and M.~Schwager, ``{Planning Periodic Persistent Monitoring Trajectories
  for Sensing Robots in Gaussian Random Fields},'' in \emph{2013 IEEE
  International Conference on Robotics and Automation}.\hskip 1em plus 0.5em
  minus 0.4em\relax IEEE, 2013, pp. 2415--2420.

\bibitem{lan2014variational}
------, ``{A Variational Approach to Trajectory Planning for Persistent
  Monitoring of Spatiotemporal Fields},'' in \emph{2014 American Control
  Conference}.\hskip 1em plus 0.5em minus 0.4em\relax IEEE, 2014, pp.
  5627--5632.

\bibitem{pinto2019monitoring}
S.~C. Pinto, S.~B. Andersson, J.~M. Hendrickx, and C.~G. Cassandras, ``{Optimal
  Multi-Agent Persistent Monitoring of the Uncertain State of a Finite Set of
  Targets},'' in \emph{Proceedings of the 2019 Control and Decision Conference
  (to appear)}.\hskip 1em plus 0.5em minus 0.4em\relax IEEE.

\bibitem{wang2019}
Y.~{Wang}, Y.~{Wei}, X.~{Liu}, N.~{Zhou}, and C.~G. {Cassandras}, ``{Optimal
  Persistent Monitoring Using Second-Order Agents With Physical Constraints},''
  \emph{IEEE Transactions on Automatic Control}, vol.~64, no.~8, pp.
  3239--3252, Aug 2019.

\bibitem{bucy1968filtering}
R.~S. Bucy and P.~D. Joseph, ``{Filtering for Stochastic Processes with
  Applications to Guidance},'' University of Southern California Los Angeles,
  Dept. of Aerospace Engineering, Tech. Rep., 1968.

\bibitem{kriegl2011denjoy}
A.~Kriegl, P.~W. Michor, and A.~Rainer, ``{Denjoy--Carleman Differentiable
  Perturbation of Polynomials and Unbounded Operators},'' \emph{Integral
  Equations and Operator Theory}, vol.~71, no.~3, p. 407, 2011.

\bibitem{lancaster1964eigenvalues}
P.~Lancaster, ``{On Eigenvalues of Matrices Dependent on a Parameter},''
  \emph{Numerische Mathematik}, vol.~6, no.~1, pp. 377--387, 1964.

\bibitem{le2010scheduling}
J.~Le~Ny, E.~Feron, and M.~A. Dahleh, ``{Scheduling Continuous-Time Kalman
  Filters},'' \emph{IEEE Transactions on Automatic Control}, vol.~56, no.~6,
  pp. 1381--1394, 2010.

\bibitem{bittanti1984periodic}
S.~Bittanti, P.~Colaneri, and G.~Guardabassi, ``{Periodic Solutions of Periodic
  Riccati Equations},'' \emph{IEEE Transactions on Automatic Control}, vol.~29,
  no.~7, pp. 665--667, 1984.

\bibitem{nicolao1992convergence}
G.~Nicolao, ``{On the Convergence to the Strong Solution of Periodic Riccati
  Equations},'' \emph{International Journal of Control}, vol.~56, no.~1, pp.
  87--97, 1992.

\bibitem{cassandras2010perturbation}
C.~G. Cassandras, Y.~Wardi, C.~G. Panayiotou, and C.~Yao, ``{Perturbation
  Analysis and Optimization of Stochastic Hybrid Systems},'' \emph{European
  Journal of Control}, vol.~16, no.~6, pp. 642--661, 2010.

\bibitem{reid1946matrix}
W.~T. Reid, ``{A Matrix Differential Equation of Riccati Type},''
  \emph{American Journal of Mathematics}, vol.~68, no.~2, pp. 237--246, 1946.

\bibitem{zhang2011matrix}
F.~Zhang, \emph{Matrix Theory: Basic Results and Techniques}.\hskip 1em plus
  0.5em minus 0.4em\relax Springer Science \& Business Media, 2011.

\bibitem{barraud1977numerical}
A.~Barraud, ``{A Numerical Algorithm to Solve AT XA-X=Q},'' in \emph{1977 IEEE
  Conference on Decision and Control including the 16th Symposium on Adaptive
  Processes and A Special Symposium on Fuzzy Set Theory and Applications},
  no.~16, 1977, pp. 420--423.

\bibitem{varga2013computational}
A.~Varga, ``{Computational Issues for Linear Periodic Systems: Paradigms,
  Algorithms, Open Problems},'' \emph{International Journal of Control},
  vol.~86, no.~7, pp. 1227--1239, 2013.

\bibitem{lin2014optimal}
X.~Lin and C.~G. Cassandras, ``{An Optimal Control Approach to the Multi-Agent
  Persistent Monitoring Problem in Two-dimensional Spaces},'' \emph{IEEE
  Transactions on Automatic Control}, vol.~60, no.~6, pp. 1659--1664, 2014.

\end{thebibliography}
\end{document}